\documentclass[fleqn,usenatbib]{mnras}

\usepackage{newtxtext,newtxmath}
\usepackage{xcolor}

\usepackage[T1]{fontenc}

\DeclareRobustCommand{\VAN}[3]{#2}
\let\VANthebibliography\thebibliography
\def\thebibliography{\DeclareRobustCommand{\VAN}[3]{##3}\VANthebibliography}

\usepackage{graphicx}	
\usepackage{amsmath}	

\newcommand{\tbn}{$\theta_{Bn}$}	

\title[Shock and turbulence]{Three-dimensional modelling of the shock-turbulence interaction}

\author[D. Trotta et al.]{D. Trotta,$^{1}$\thanks{E-mail: d.trotta@imperial.ac.uk }
O. Pezzi,$^{2}$
D. Burgess,$^{3}$
L. Preisser,$^{4}$
X. Blanco-Cano,$^{5}$
P. Kajdic,$^{5}$
H. Hietala,$^{3}$
\newauthor
T. S. Horbury,$^{1}$
R. Vainio,$^{6}$
N. Dresing,$^{6}$
A. Retin\`{o},$^{7}$
M. F. Marcucci,$^{8}$
L. Sorriso-Valvo,$^{2,9,10}$
S. Servidio$^{11}$
\newauthor
and F. Valentini$^{11}$
\\
$^{1}$ The Blackett Laboratory, Department of Physics, Imperial College London, London SW7 2AZ, UK \\
$^{2}$ Istituto per la Scienza e Tecnologia dei Plasmi (ISTP), Consiglio Nazionale delle Ricerche, Via Amendola 122/D, I-70126 Bari, Italy \\
$^{3}$ Department of Physics and Astronomy, Queen Mary University of London, E1 4NS, UK \\
$^{4}$ Space Research Institute, Austrian Academy of Sciences, Graz, Austria \\
$^{5}$Departamento de Ciencias Espaciales, Instituto de Geofísica, Universidad Nacional Autónoma de México, Ciudad Universitaria, Ciudad de México, Mexico \\
$^{6}$Department of Physics and Astronomy, University of Turku, FI-20014 Turku, Finland \\
$^{7}$LPP, Ecole Polytechnique, CNRS, UPMC, Université Paris Sud, Palaiseau, France\\
$^{8}$INAF-Istituto di Astrofisica e Planetologia Spaziali, Rome, Italy \\
$^{9}$Space and Plasma Physics, School of Electrical Engineering and Computer Science, KTH Royal Institute of Technology, Stockholm, Sweden \\
$^{10}$Swedish Institute of Space Physics (IRF), \r{A}ngstr\"om Laboratory, L\"agerhyddsv\"agen 1, 75121 Uppsala, Sweden \\
$^{11}$Dipartimento di Fisica, Universit\`a della Calabria, Rende 87036, Italy \\
}

\date{Accepted XXX. Received YYY; in original form ZZZ}

\pubyear{2023}

\begin{document}
\label{firstpage}
\pagerange{\pageref{firstpage}--\pageref{lastpage}}
\maketitle

\begin{abstract}
The complex interaction between shocks and plasma turbulence is extremely important to address crucial features of energy conversion in a broad range of astrophysical systems. We study the interaction between a supercritical, perpendicular shock and pre-existing, fully-developed plasma turbulence, employing a novel combination of magnetohydrodynamic (MHD) and small-scale, hybrid-kinetic simulations where a shock is propagating through a turbulent medium. The variability of the shock front in the unperturbed case and for two levels of upstream fluctuations is addressed. We find that the behaviour of shock ripples, i.e., shock surface fluctuations with short (a few ion skin depths, $d_i$) wavelengths, is modified by the presence of pre-existing turbulence, which also induces strong corrugations of the shock front at larger scales. We link this complex behaviour of the shock front and the shock downstream structuring with the proton temperature anisotropies produced in the shock-turbulence system. Finally, we put our modelling effort in the context of spacecraft observations, elucidating the role of novel cross-scale, multi-spacecraft measurements in resolving shock front irregularities at different scales. These results are relevant for a broad range of astrophysical systems characterised by the presence of shock waves interacting with plasma turbulence.

\end{abstract}




\section{Introduction}
\label{sec:introduction}

Collisionless shocks are fundamental components of our universe, crucial in reconstructing the properties of a broad range of astrophysical environments~\citep{Amato2017,Brunetti2014}. Generally speaking, shock waves convert directed flow energy (upstream) into heat and magnetic energy (downstream). In the collisionless case, a fraction of the available energy can be channeled into the production of energetic particles, a pivotal feature to understand many aspects of \textit{in-situ} and remote observations~\citep{Burgess2015}. Thus, collisionless shocks play a fundamental role in energy conversion in a variety of systems, ranging from solar flares~\citep{Woo1981} to interacting galaxy clusters~\citep{Bykov2019}. While some aspects of energy conversion at shock waves are not fully understood despite decades of research, a picture invoking a complex shock behaviour is emerging~\citep[e.g.,][]{Treumann2009}. 

One of the most important parameters controlling shock structure and behaviour is the shock normal angle, i.e., the angle between the normal to the shock surface and the upstream magnetic field, \tbn. Shocks with \tbn $ \lesssim 45^\circ$ (i.e., for which the upstream magnetic field and the shock normal are well-aligned) are called quasi-parallel, while in the quasi-perpendicular case \tbn $ \gtrsim 45^\circ$. Other important parameters are the shock Alfv\'enic and sonic Mach numbers, defined as $\rm M_A = v_{sh}/v_A$ and $\rm M_S = v_{sh}/c_s$, respectively, and the plasma $\beta = \rm v_{th}^2/v_A^2$. Here,  $\rm v_{sh}$ is the shock speed in the upstream flow frame, while $\rm v_A$, $\rm c_s$ and $\rm v_{th}$ are the Alfv\'en, sound and thermal speed in the region upstream from the shock.

\begin{figure*}
	\includegraphics[width=\textwidth]{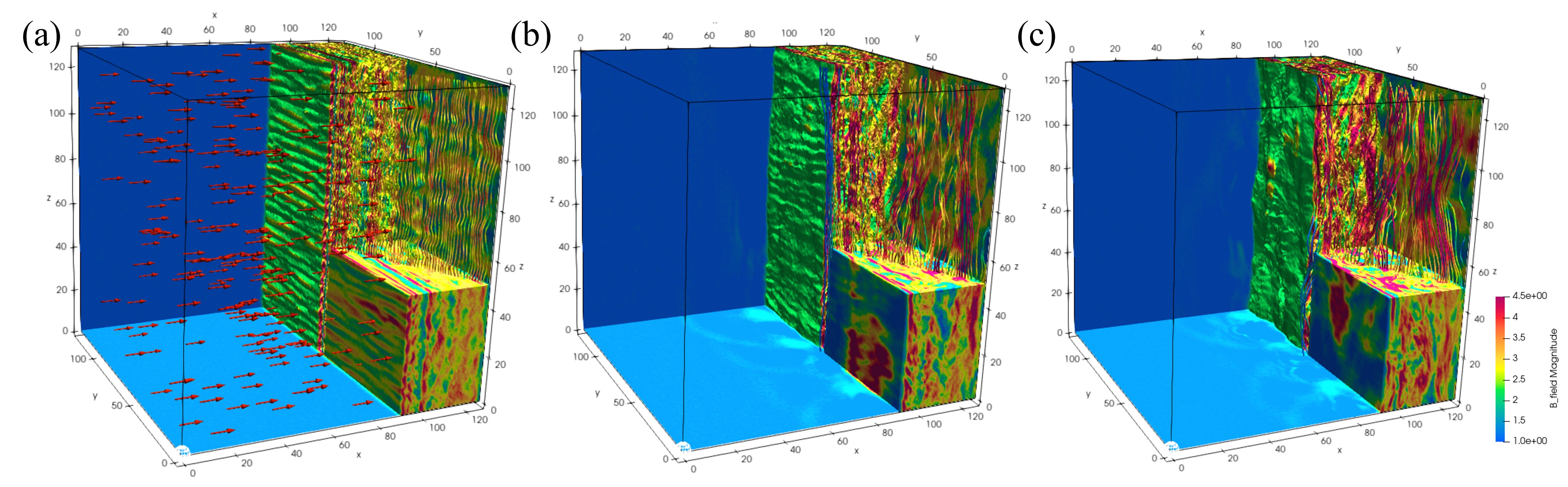}
    \caption{Overview of the simulations presented in this work, ordered for increasing level of perturbation $\delta B/B_0 = 0, 0.5$ and $1$ ((a), (b) and (c), respectively). In each volume, we plot the magnetic field magnitude for the $z=0$, $x=256$ and $y=256$ planes. An isocontour for $B>2 B_0$ is shown in a subvolume, rendering the shock surface. Slices on the shock front and downstream are also shown together with some magnetic field lines integrated downstream. For the unperturbed case (a), streamlines of upstream bulk flow speed are also shown for reference (red arrows). All three renderings are done at simulation time $\rm T\Omega_{ci} = 16$.}
    \label{fig:fig1_overview}
\end{figure*}

Shocks in the heliosphere are unique because they are accessible by \textit{in-situ} spacecraft exploration \citep{Richter1985}, thus providing the missing link to the remote observations of astrophysical systems. In this picture, the Earth's bow shock, resulting from the interaction between the supersonic solar wind and the Earth's magnetosphere, has become the most studied shock using direct observations \citep{Formisano1979}. More recently, the Magnetospheric MultiScale mission \citep[MMS,][]{Burch2016}, elucidated novel aspects of the overall energetics of the shock system \citep{Schwartz2022}. Other heliospheric shocks that can be observed \textit{in-situ} are interplanetary shocks, consequence of solar activity such as, for example, coronal mass ejections~\citep{Kilpua2015,BlancoCano2016}. Such studies highlight the importance of various kinds of shock irregularities for understanding how plasma is processed across a shock wave~\citep[e.g.,][]{Lobzin2007,Wilson2009, Kajdic2019, Trotta2023a}.

A particularly interesting kind of shock irregularity is shock rippling, i.e., surface fluctuations, recently observed \textit{in-situ} with MMS at the quasi-perpendicular Earth's bow shock~\citep{Johlander2016}. We distinguish this small-scale rippling from larger scale perturbations of the shock front due to self-generated upstream waves being advected back at the shock, also important, especially at geometries departing from the perpendicular one~\citep[see, for example][]{Kajdic2019,Turc2023}. Shock rippling in quasi-perpendicular geometries happens at supercritical (i.e., $\rm M_A \gtrsim 3$) shocks, where ion reflection at the shock front leads to the foot-ramp-overshoot structure~\citep[see][]{Kivelson1995}.  Such structuring is characterised by highly anisotropic, non-thermal particle distributions in the foot and ramp, due to the presence of incident and reflected populations, often particularly challenging to observe in-situ. Earlier theoretical and numerical studies elucidated that the non-thermal distributions in the shock foot and ramp lead to shock ripples that have short wavelength (about a few ion skin depths) propagate along the shock front at the Alfv\'en speed of the overshoot~\citep{Lowe2003,Burgess20163d}. Shock rippling was also proven to be crucial for efficient electron acceleration at shocks in a variety of astrophysical environments~\citep{Trotta2019,Kang2019, Kobzar2021}.  

Another important feature of quasi-perpendicular shocks, consequence of the behaviour discussed above, is the presence of a strong perpendicular temperature anisotropy, routinely observed downstream of the quasi-perpendicular bow shock of Earth~\citep{Eastwood2015}. The small-scale pattern of the temperature anisotropy typical of quasi-perpendicular shocks has also been investigated using numerical simulations \citep{Burgess2007,Preisser2020, Ofman2021}. Numerical modelling is invaluable for understanding details of the shock dynamics that are often challenging to observe~\citep[e.g.,][]{Krasnoselskikh2002, Caprioli2014, Matsumoto2015, Gedalin2018}. 

An ubiquitous property of our universe is plasma turbulence~\citep[e.g.,][]{Lazarian2012}, crucial for energy dissipation in collisionless plasmas~\citep[e.g.,][]{Matthaeus2015,Matthaeus2020,Pezzi2021a}. Turbulence is also a fundamental phenomenon leading to particle acceleration, as shown by Fermi's early works~\citep{Fermi1949, Fermi1954} and in decades of subsequent research~\citep[e.g.,][]{Vlahos2004, Kowal2012, Guo2021b} (see also \citep{Khabarova2021,Pezzi2021b} for a review). The shock -- turbulence interaction is an important and often spectacular pathway to efficient energy conversion and particle acceleration~\citep{Zank2002, Guo2021}, and the transport properties of shock accelerated particles have been shown to depend on the level of upstream fluctuations~\citep{Lario2022}. Numerical simulations are particularly useful in addressing the complex interaction between shock waves and (pre-existing) plasma turbulence. Early efforts modelling shock waves propagating in an upstream medium perturbed with a prescribed set of fluctuations have shown that both the shock front behaviour and the production of energetic particles are influenced by the upstream conditions~\citep{Giacalone2005a, Guo2015}. The behaviour of energetic particles across turbulence-mediated shocks, and shocks interacting with trains of current sheets was also investigated by \citet{Nakanotani2021, Nakanotani2022}, revealing enhanced particle energisation due to turbulence. Recently, \citet{Trotta2021} looked at the interaction between fully-developed turbulence and oblique shocks in two dimensions, finding enhanced particle transport in phase space in such an interaction, with pre-existing turbulence providing a source of strong upstream scattering for the shock-reflected particles. The important problem of how turbulent structures are transmitted across shock waves was also investigated with a combination of simulations and Earth's bow shock observations, revealing a magnetic helicity increase due to turbulent structures' compression at the shock \citep{Trotta2022a}.

In this work, we address, in fully three-dimensional geometry, the interaction of a rippled, perpendicular shock front with fully-developed upstream turbulence. To this end, we employ a combination of MHD and small-scale, kinetic simulations with different pre-existing, upstream turbulence strength. The shock front dynamics are addressed, revealing a complex interplay in which ripples may survive or get inhibited due to local perturbations. The temperature anisotropy across the shock transition is also studied, to see how the scenario in which a strong anisotropy generated at the shock ramp relaxes towards equilibrium downstream of the shock is modified by turbulent fluctuations. Finally, we show how a multi-scale, multi-spacecraft approach is needed to properly address the properties of the shock-turbulence system, in support for future missions such as HelioSwarm~\citep{Spence2019} and Plasma Observatory, a space mission proposal candidate to the next M7 call of the European Space Agency~\citep{Retino2022}. The paper is organised as follows: in Section~\ref{sec:data_methods} the simulation methods are described; in Section~\ref{sec:results_disc} the results are presented and discussed, and Section~\ref{sec:conclusions} contains the conclusions of the work.

\section{Methods}
\label{sec:data_methods}
\begin{figure}	\includegraphics[width=.48\textwidth]{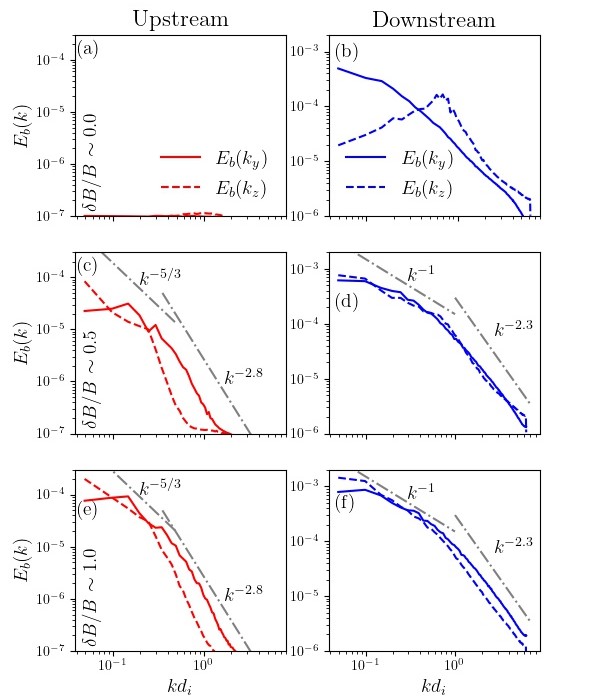}
    \caption{One-dimensional (reduced) magnetic field spectra during the shock-turbulence interaction, computed along the $z$-direction of the mean magnetic field and the $y$-direction, perpendicular to both the mean magnetic field and the shock normal (dashed and continuous line, respectively). The spectra are averaged in the shock upstream and downstream (left to right) for all the simulations (top to bottom), respectively in the regions $x\in[75,90] d_i$ and $x\in[105,120]d_i$. The grey dotted-dashed lines show examples of power-law scaling 
 relevant to turbulent spectra.}
    \label{fig:fig2_spec}
\end{figure}

\begin{figure*}
	\includegraphics[width=\textwidth]{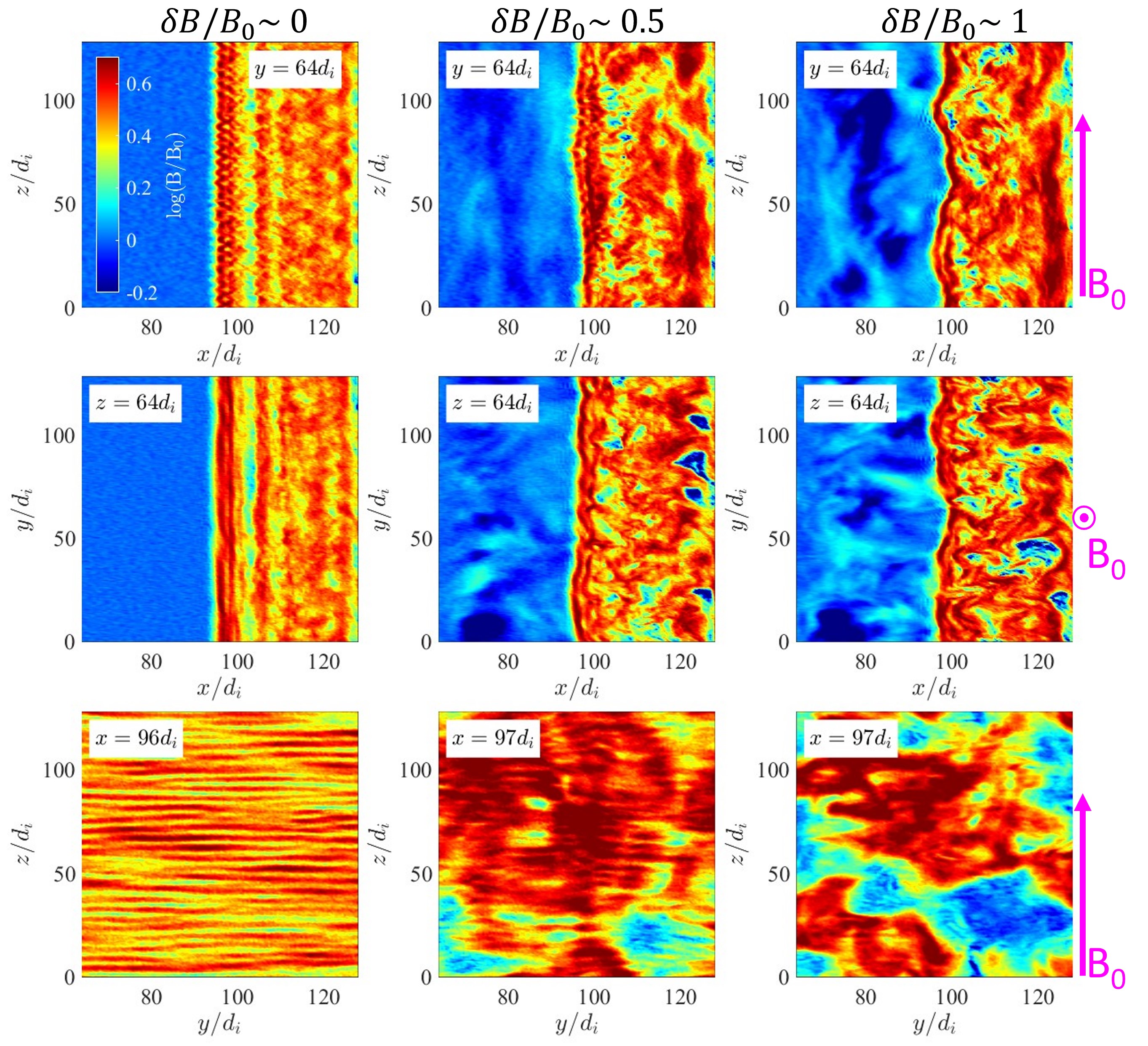}
    \caption{Two-dimensional slices of the magnetic field magnitude for the $x-z$, $x-y$ and $y-z$ (top to bottom) for the three simulations at time $\rm T \Omega_{ci} = 14$, organised with increasing level of turbulent fluctuations (left to right). The magenta arrows on the right-hand side of the figure display the mean magnetic field direction.}
    \label{fig:fig3_fronts}
\end{figure*}

\begin{figure*}
	\includegraphics[width=\textwidth]{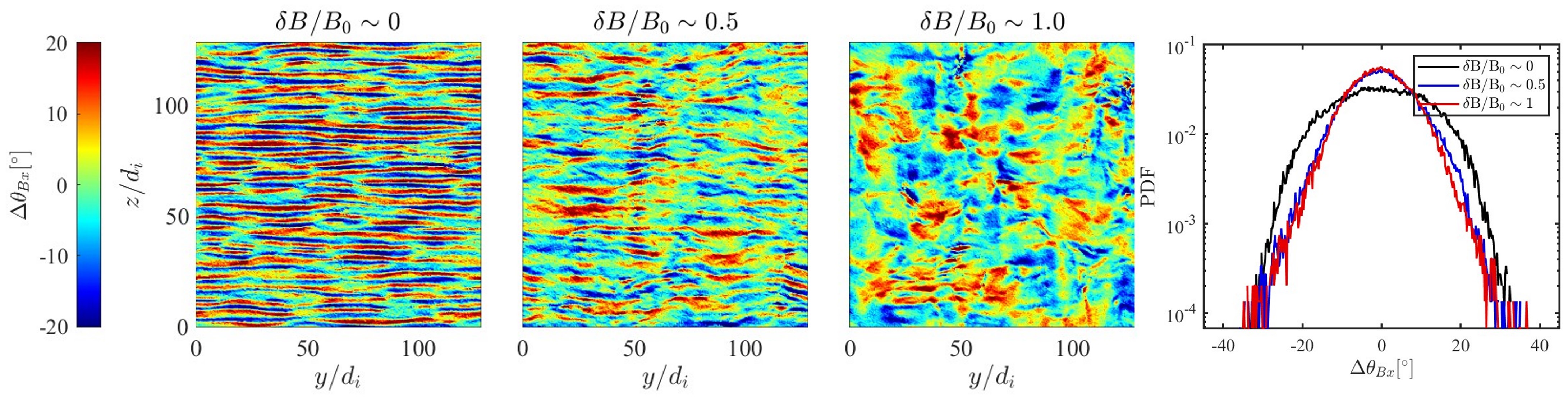}
    \caption{Departure from the nominal shock normal angle $\Delta \theta_{Bx}$ along the shock front in all simulation cases, at time $\rm T \Omega_{ci} = 14$, for increasing level of turbulence strength (left to right). A PDF of such values is shown in the right panel.}
    \label{fig:fig4_tbn}
\end{figure*}

Our numerical simulations are carried out in two stages, as done in reduced, two-dimensional geometry in \citet{Trotta2021,Trotta2022a}. In order to inspect the interaction of shock waves with the coherent structures of turbulence, first, MHD simulations are used to produce turbulent fields, which are then used in the second (main) stage of the simulations to perturb the initial condition of a hybrid Particle-In-Cell (PIC) shock simulation, obtaining a shock that propagates in a turbulent upstream plasma.

Three dimensional, compressible MHD simulations are used to generate fully-developed, decaying turbulence. To this purpose, a pseudo-spectral algorithm that adopts second-order Runge-Kutta scheme to advance in time the MHD equations was used. Such a code was recently extended to the fully three-dimensional configuration starting from a previous two-dimensional algorithm \citep{Vasconez2015}, already adopted to investigate, for example, the interaction of two counterpropagating Alfv\'enic wavepackets \citep{Pezzi2017a,Pezzi2017b}, and the parametric instability \citep{Primavera2019}.

Two simulations of turbulence were performed, initialised with different levels of turbulence fluctuations, $\delta B/B_0 = 0.5, 1.0$, where ${\bf {B}_0} = B_0 {\bf\hat{z}} $ is the mean field, and $\delta B$ is the rms level of the fluctuations. At the time instant in which turbulence is most intense in the MHD simulation, i.e. when $\langle|{\bf{j}}|^2\rangle$ reaches its maximum value being $\bf{j}=\nabla \times \bf{B}$, the output is stored to be used as an initial condition for the shock simulation, where magnetic field and ion bulk flow speed are perturbed, as done in two-dimensions in \citet{Trotta2021}. 
In the MHD simulations, standard normalization has been adopted: time, space, and velocities are respectively scaled to the Alfv\'en time $t_A$, a generic length $L_A$, and the Alfv\'en speed $\mathrm{v_A}=L_A/t_A$. The tri-periodic cubic box, of size $L_0=2\pi L_A$, has been discretised with $256$ gridpoints along each direction.

Shock simulations with perturbed and unperturbed (laminar) upstream conditions are then performed using the HYPSI code~\citep[e.g.,][]{Trotta2020a}. Here, protons are modelled as macroparticles and advanced using the standard PIC method~\citep{Birdsall1991}. The electrons, on the other hand, are modelled as a massless, charge-neutralizing fluid with an adiabatic equation of state. The HYPSI code is based on the Current Advance Method and Cyclic Leapfrog (CAM-CL) algorithm \citep[][]{Matthews1994}. The shock is initiated by the injection method \citep[][]{Quest1985}, in which the plasma flows in the $x$-direction with a defined (super-Alfv\'enic) velocity $V_\mathrm{in}$. The right-hand boundary of the simulation domain acts as a reflecting wall, and at the left-hand boundary plasma is continuously injected. The simulation is periodic in the $y$- and $z$-directions. A shock is created as a consequence of reflection at the wall, and propagates in the negative $x$-direction. In the simulation frame, the (mean) upstream flow is along the shock normal. To ensure that the $\nabla \cdot \mathbf{B} = 0$ equation is satisfied in the non-periodic shock simulations, the perturbations go to zero a the simulation boundaries, and the perturbation introduced is therefore limited in space and time, due to the fact that freshly injected plasma at the left-hand side of the simulation is unperturbed.

In the hybrid simulations, distance is normalised to the ion inertial length $d_i \equiv c/\omega_{pi}$, time to the inverse cyclotron frequency ${\Omega_{ci}}^{-1}$, velocity to the Alfv\'en speed $\rm v_A$ (all referred to the unperturbed upstream state), and magnetic field and density to their unperturbed upstream values, $B_0$ and $n_0$, respectively. The nominal angle between the shock normal and the upstream magnetic field, $\theta_{Bn}$, is 90$^\circ$, with the upstream magnetic field along the $z$-direction. We set the upstream flow speed to $V_\mathrm{in} = 4.5 \rm v_A$, and the resulting Alfv\'enic Mach number of the shock is approximately $M_A \sim 6$. The upstream ion distribution function is an isotropic Maxwellian and the ion $\beta_i$ is 1 (typical of solar wind plasma~\citep{Wilson2018}). The simulation $x-y-z$ domain  is 128 $\times$ 128 $\times$ 128 $d_i^3$. The spatial resolution used is $\Delta x$ = $\Delta y$= $\Delta z$ = 0.5 $d_i$. The final time for the simulation is 20 $\Omega_{ci}^{-1}$, and the time step for particle {(ion)} advance is $\Delta t_{}$  = 0.01 $\Omega_{ci}^{-1}$. Substepping is used for the magnetic field advance, with an effective time step of $\Delta t_{B} = \Delta t_{}/10$. A small, nonzero  resistivity is introduced in the magnetic induction equation and its value is set so that there are not excessive fluctuations at the grid scale. The number of particles per cell used is always greater than 50 (upstream).

Three simulations are presented in this work, with the same nominal shock parameters in the unperturbed case $\delta B/B_0 \sim 0$ together with the two perturbed cases $\delta B/B_0 = 0.5$ and $ 1.0$.

\section{Results and discussion}
\label{sec:results_disc}

\subsection{Perturbed shocks simulations overview}
\label{subsec:sim_oview}

In Figure~\ref{fig:fig1_overview}, we present an overview of our simulations, showing three snapshots taken during the shock-turbulence interaction. In the magnetic field rendering for the unperturbed case (Figure~\ref{fig:fig1_overview}(a)), it is possible to see the rippled shock front, a result compatible with previous simulations of perpendicular shocks interacting with a laminar upstream flow~\citep[e.g.,][]{Burgess20163d}. In this case, the downstream region also reveals shock-induced fluctuations, with the overshoot -- undershoot structure typical of supercritical shocks being visible immediately behind the shock.

The presence of upstream turbulence induces strong modifications with respect to the unperturbed case. As can be seen in Figure~\ref{fig:fig1_overview}(b) and (c), the shock front appears distorted in the presence of upstream turbulence, due to convection of the fluctuations through the shock front. A more complex downstream scenario is also observed. Interestingly, in the moderately perturbed case, shock rippling survives the presence of upstream turbulence, and keeps operating at the distorted shock front. Finally, in the strongly perturbed case, the interplay between ripples and shock distortions due to turbulence becomes even more complex.

We further characterise the turbulent shock environments with the magnetic field spectral density in the shock upstream and downstream, for all cases. Figure~\ref{fig:fig2_spec} shows one-dimensional magnetic field spectra for the $z$- and $y$-directions, parallel and perpendicular to the mean magnetic field, respectively. For all the simulations, the spectra have been computed by using Fast Fourier Transforms. One-dimensional spectra along the $z$-direction are computed averaging over the $y$ direction (and vice versa for the spectrum in the $y$-direction). Further averaging is performed along the nominal shock normal ($x$) direction. To this end, the upstream and downstream regions have been defined by the conditions $75 \, d_i < x < 90 \, d_i$ and  $105 \, d_i < x < 120 \, d_i$, respectively, at simulation time $\rm T \Omega_{ci} = 14$, when the average shock position is of 95~$d_i$. Therefore, the shock front highlighted in Figure~\ref{fig:fig1_overview} is excluded from this diagnostic, focusing on the effect of the shock passage in the processing of turbulence. 

In the unperturbed case (panels $\rm a$ and $\rm b$ of Figure~\ref{fig:fig2_spec}), a spectrum of downstream fluctuations (blue lines) develops due to the shock passage, with a strong injection in the parallel spectrum that shows an energy bump at $k_z d_i \sim 1$ associated with the ripples propagating parallel to the shock front and along the mean magnetic field (Figure~\ref{fig:fig2_spec}(b)).
The other panels of Figure~\ref{fig:fig2_spec} show how turbulence is affected by the shock crossing, with two major effects: (i) the increase of the level of turbulent fluctuations, and (ii) the isotropisation of turbulent energy. 
The upstream spectra (panels $c$ and $e$, red lines) show anisotropies in the $k_y$ -- $k_z$ directions, a well-known feature of MHD turbulence~\citep[][]{Shebalin1983}. 
In both cases, the perpendicular spectrum presents a short Kolmogorov-like scaling ($k^{-5/3}$) at small wavevectors, this being limited by the dynamical range of underlying MHD simulations, while the parallel spectrum has smaller power and nearly no power-law scaling. 
At sub-ion scales, both spectra are steeper ($\sim k^{-2.8}$), indicating energy dispersion and dissipation. In panels $\rm c$--$\rm f$, the grey lines indicate typical plasma turbulence power-laws \citep[e.g.][]{chen_2016}, shown for reference.
This spectral behaviour is compatible with typical solar wind turbulence observations~\citep[e.g.][]{Chen2014} and previous kinetic simulations~\cite[e.g.,][]{Perrone2013, Franci2018}.
The analysis of the downstream spectra (panels $d$ and $f$, blue lines) shows that the overall level of fluctuations increases due to the shock compression (notice the different range in $y$-axis of left and right panels in Fig. \ref{fig:fig2_spec})~\citep{Pitna2017, Zhao2021}. 
The downstream spectra show a behaviour compatible with observations of turbulence in the terrestrial magnetosheath~\citep{Huang2017}, with the absence of a Kolmogorov scaling, replaced by an energy-containing $k^{-1}$ range, followed by a transition to a marked steepening at sub-ion scales~\citep{Sahraoui2020}. 
The spectral anisotropy is greatly reduced (Figure~\ref{fig:fig2_spec}(d,f)), in particular for the intermediate case of turbulence strength ($\delta B/B_0 \sim 0.5$). 
This may be due either to an isotropisation effect induced by the shock crossing or to the interplay between pre-existing fluctuations and shock-induced fluctuations. 

\begin{figure*}
	\includegraphics[width=\textwidth]{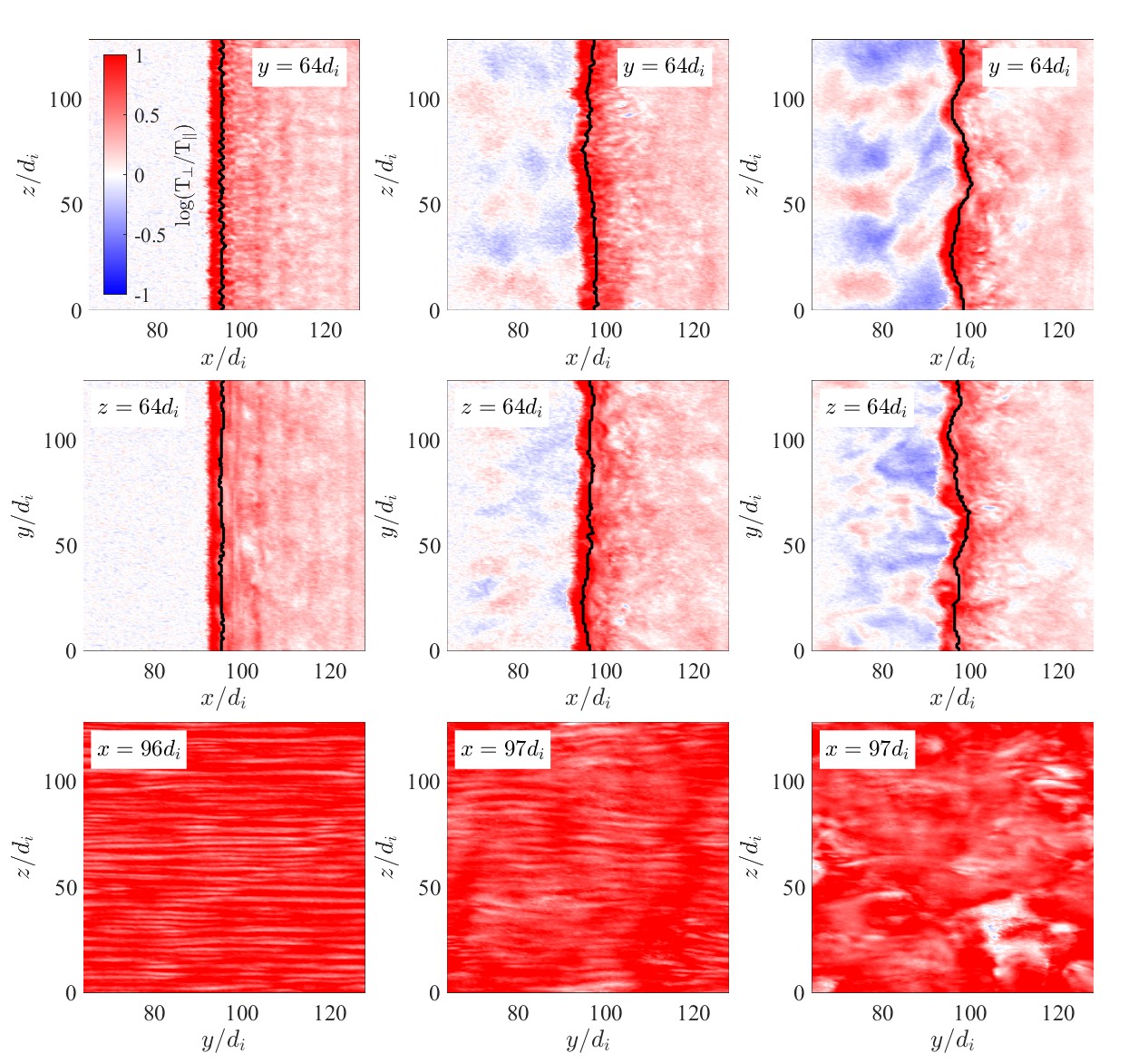}
    \caption{Two dimensional slices of the proton temperature anisotropy $\mathrm{log(T_{\perp}/T_{||})}$, or the $x-z$, $x-y$ and $y-z$ (top to bottom) for the three simulations at time $\rm T \Omega_{ci} = 14$, organised with increasing level of turbulent fluctuations from left to right (as done in Figure~\ref{fig:fig3_fronts} for the magnetic field). The black line shows the shock front position.}
    \label{fig:fig5_tanis_fronts}
\end{figure*}

\subsection{Shock front behaviour}
\label{subsec:fronts}

In this section, we discuss the details of the observed shock front behaviour. Figure~\ref{fig:fig3_fronts} shows two-dimensional slices of the shock-turbulence interaction simulations. The shock rippling is the predominant feature of the unperturbed shock front, with magnetic field fluctuations along the shock front showing at typical spatial scales of some $d_i$. The ripples propagate along the shock front in the mean magnetic field direction, as elucidated in detail in \citet{Burgess20163d}. In the top left panel of Figure~\ref{fig:fig3_fronts}, it is possible to appreciate how shock rippling participates in the shock overshoot--undershoot structuring, namely as a rapidly fluctuating feature visible in the plane containing the mean magnetic field, superimposed to the large scale structuring observed from the shock front and in the downstream region.

The upstream turbulence has the major effect of introducing shock front irregularities at the scales where the turbulent cascade is operating, clearly seen as shock front undulations in the perturbed cases, happening at larger scales than the self-induced shock rippling. It is important to note that the shock front irregularities introduced by the turbulence do not depend on the shock front behaviour. This represents a fundamental difference with respect to other cases where shock front corrugation is observed as a result of self-generated upstream waves/fluctuations, also generating shock front distorsion~\citep[see][]{Kajdic2021,Turc2023}. As hinted in the discussion above, small-scale shock rippling is clearly present in the moderately perturbed case, as it can be seen for the shock front in the $\delta B/B_0 \sim 0.5$ case. However, due to the changes in the mean magnetic field at turbulent fluctuation scales, their propagation becomes more complex along the shock front, in a scenario in which different ``patches'' of the shock front have ripples with different orientations, with potential implications for efficient particle acceleration. Furthermore, while ripples survive at the shock front, we note that the region downstream of the shock becomes much more complex than in the laminar case, due to the variability introduced by the turbulent fluctuations and the irregularity in the shock front. Finally, such complexity is further enhanced in the strongly perturbed case, where a highly dynamic shock front is observed. The signature of shock rippling becomes increasingly hard to disentangle with respect to other irregularities at play in the shock front. 

We also note that, due to turbulent structures being transmitted from upstream to downstream, the perturbed cases allow for larger amplitude depletions in magnetic field magnitude downstream, a feature consistent with studies carried out in reduced, two-dimensional geometry~\citep{Nakanotani2022}. The three-dimensional behaviour of the transmitted turbulent structures and their importance as extra sources of energetic particles beyond energisation at the shock front through turbulent acceleration mechanisms~\citep{Drake2006,Comisso2022} is an extremely interesting topic, which will be the subject of further investigation.

We further investigate the shock front behaviour by analysing the departures from the expected shock normal angle along the shock front. Given the simulation setup (see Section~\ref{sec:data_methods}), the nominal shock $\theta_{Bn}$ for the shocks simulated here is $\theta_{Bn} = \theta_{Bx} = 90^\circ$. Due to its three-dimensional structure, particularly important in the perturbed cases, we define the shock front position as where $B > 3 B_0$ (following a similar procedure presented in \citet{Kajdic2019}). The local shock normal angle is then computed as $\theta_{Bx}(y,z) = \mathrm{cos^{-1}}(B_x(y,z)/B(y,z))$. 

Such analysis is shown in Figure~\ref{fig:fig4_tbn}, where the local values of $\theta_{Bx}$ are displayed. These change rapidly in the unperturbed case, with even strong departures (up to about 40$^\circ$ from the nominal value, due to shock rippling, consistent with what previously shown in \citet{Trotta2019}. When upstream turbulence is included, the picture significantly changes. Departures from the nominal shock geometry happen over a wider range of spatial scales, introduced by the turbulence, with important implications on the interplay between the shock and its surroundings. In the turbulent cases, the small-scale ripples appear to induce weaker changes in the local shock geometry at short wavelengths (see the Probability Density Function in Figure~\ref{fig:fig4_tbn}), due to the upstream mixing introduced by the turbulence. This result is extremely interesting and relevant when addressing the dynamics of upstream particles interacting with different portions of the shock front showing different local geometries in a variety of scales.

\subsection{Temperature anisotropies}
\label{subsec:temp_anis}

Shock rippling is a consequence of the perpendicular temperature anisotropy driven in the shock foot by the reflected protons~\citep{Winske1988}. It is therefore natural to study such temperature anisotropies in the simulations, addressing their relation with the observed shock irregularities.

Such analysis is carried out in Figure~\ref{fig:fig5_tanis_fronts}, where two-dimensional slices of the simulation domain for the quantity $\mathrm{ log(T_{\perp}/T_{||})}$ are shown, in the same format and time as Figure~\ref{fig:fig3_fronts} for the magnetic field. The shock front position has been calculated with the same criterion used for Figure~\ref{fig:fig4_tbn} (see Section~\ref{subsec:fronts}). Here, the parallel and perpendicular temperatures have been computed by projecting the proton temperature tensor along the local magnetic field in the simulations. In the unperturbed case, the typical scenario for the supercritical perpendicular shock is recovered, with the presence of a strong perpendicular temperature anisotropy ($\rm T_{\perp}/T_{||} > 1$) at the shock front (see the left panels of Figure~\ref{fig:fig5_tanis_fronts}), relaxing in the downstream region. It is possible to identify oscillations in the temperature anisotropies, happening at wavelengths that increase with the distance from the shock~\citep{Lu2006, Preisser2020}. We note that far downstream of the shock, plasma has not yet relaxed to an isotropic configuration, due to the limited size of the simulation domain. However, the main focus of this study is the shock front behaviour in response to upstream turbulence, and therefore the interesting study of asymptotic behaviour of the temperature anisotropy, and the associated instabilities~\citep{Hellinger2006,Kim2021} in presence of pre-existing turbulence is beyond scope.

When the shock propagates through turbulent media, many interesting features arise. It can be seen that the shock does not propagate anymore in an isotropic medium. Along the (distorted) shock front, a strong perpendicular temperature anisotropy is found, but the structuring seen in the unperturbed case is modified by the turbulent fluctuations, as it can be seen, for example, in the $\delta B/B_0 \sim 0.5$ case. The pre-existing fluctuations, together with the strongly distorted shock geometry allow for regions of parallel temperature anisotropy along the shock front, upstream of it and in the close downstream region (see the right-hand panels of Figure~\ref{fig:fig5_tanis_fronts}), an important aspect of the shock -- turbulence interaction. Such a complexity in temperature anisotropy explains the modified rippling found in the magnetic field analysis.

Another crucial feature emerging from Figure~\ref{fig:fig5_tanis_fronts} is the difference in the shock downstream regions for increasing levels of turbulence. In particular, comparing the $\delta B/B_0 \sim 0$ and $\delta B/B_0 \sim 1$ cases, we find that the shock downstream region in the strongly perturbed case appears more ``isotropic'' than the unperturbed case, that is, large regions of temperature isotropy are found downstream of the strongly perturbed shock.

\begin{figure}
	\includegraphics[width=.45\textwidth]{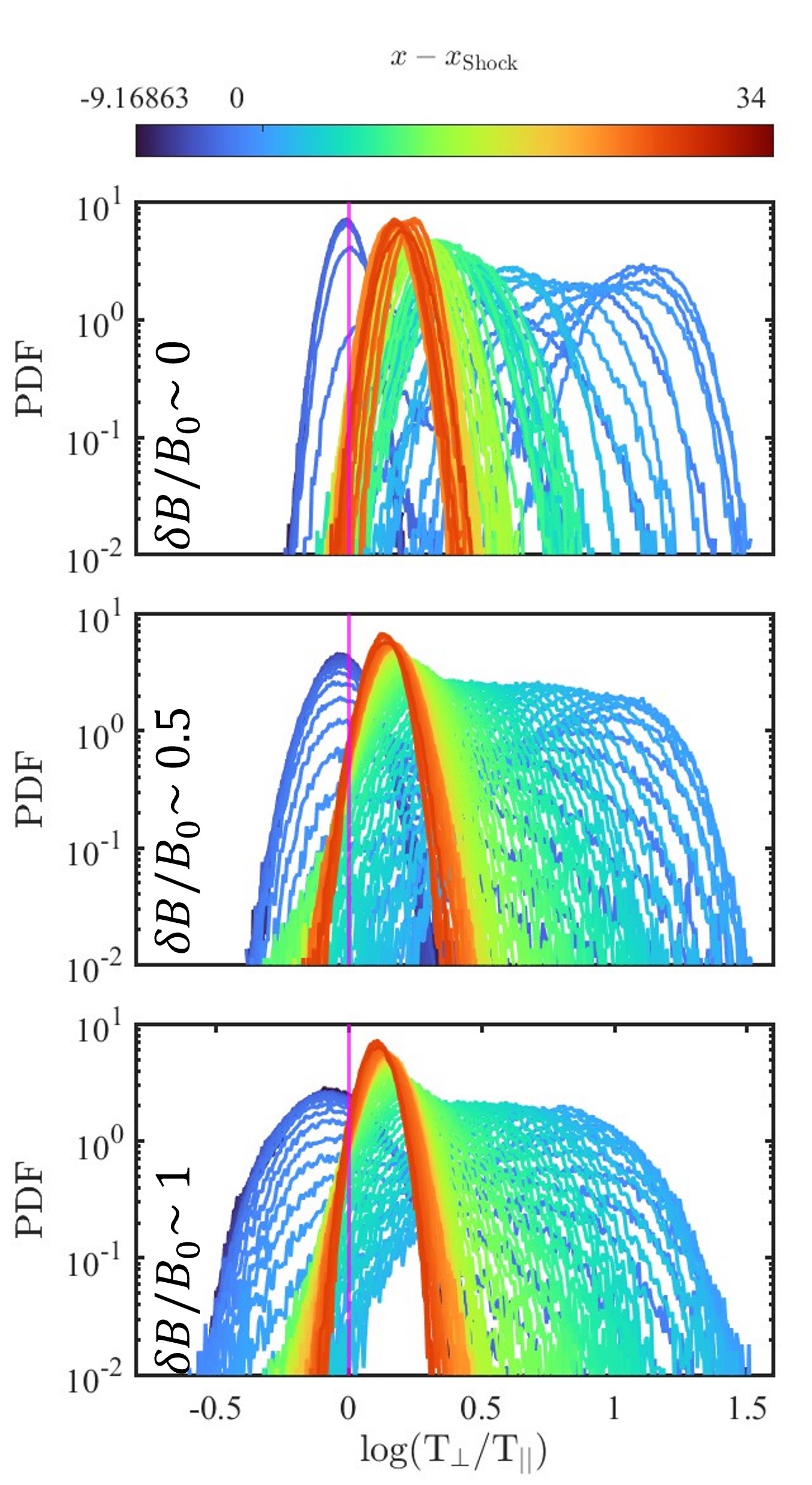}
    \caption{Temperature anisotropy PDFs, computed in $y-z$ planes at simulation time $\rm T \Omega_{ci} = 14$ (as in Figure~\ref{fig:fig5_tanis_fronts}) for different distances from the shock front (colors) for cases with increasing level of upstream turbulence (top to bottom). The vertical magenta line marks the isotropic configuration $\mathrm{T_{\perp}/T_{||} = 1}$.}
    \label{fig:fig6_tanis_paths}
\end{figure}

To make this point more quantitative, we studied the PDF of the temperature anisotropy in $y-z$ planes (parallel to the shock front) as a function of the distance from the shock in the three cases, shown in  Figure~\ref{fig:fig6_tanis_paths}. Here, PDFs with different colors are collcted at different distances from the shock (which is at zero), while the vertical magenta line indicates $\rm T_{\perp}/T_{||} = 1$. Many interesting features are revealed by this analysis. First of all, the largest values for the perpendicular temperature anisotropy are achieved in the unperturbed case and in the vicinity of the shock front (top panel of Figure~\ref{fig:fig6_tanis_paths}). Then, due to the increasing turbulence strength, in the most turbulent case the PDFs are closest to isotropy downstream, as hinted in the discussion above. Thus, when pre-existing turbulence is strong, the out-of equilibrium configurations induced by the shock front are decay faster (i.e., closer to the shock front) with respect to a laminar upstream plasma. Finally, it may be also noted that for stronger turbulence, configurations of parallel temperature isotropy become increasingly probable, due to the pre-existing population of fluctuations being transmitted across the shock front and also due to the strong local geometry changes induced by the turbulent fluctuations. Consequently, the probability of having, locally, populations of backstreaming ions becomes larger for larger upstream turbulent strength.

\begin{figure*}
	\includegraphics[width=\textwidth]{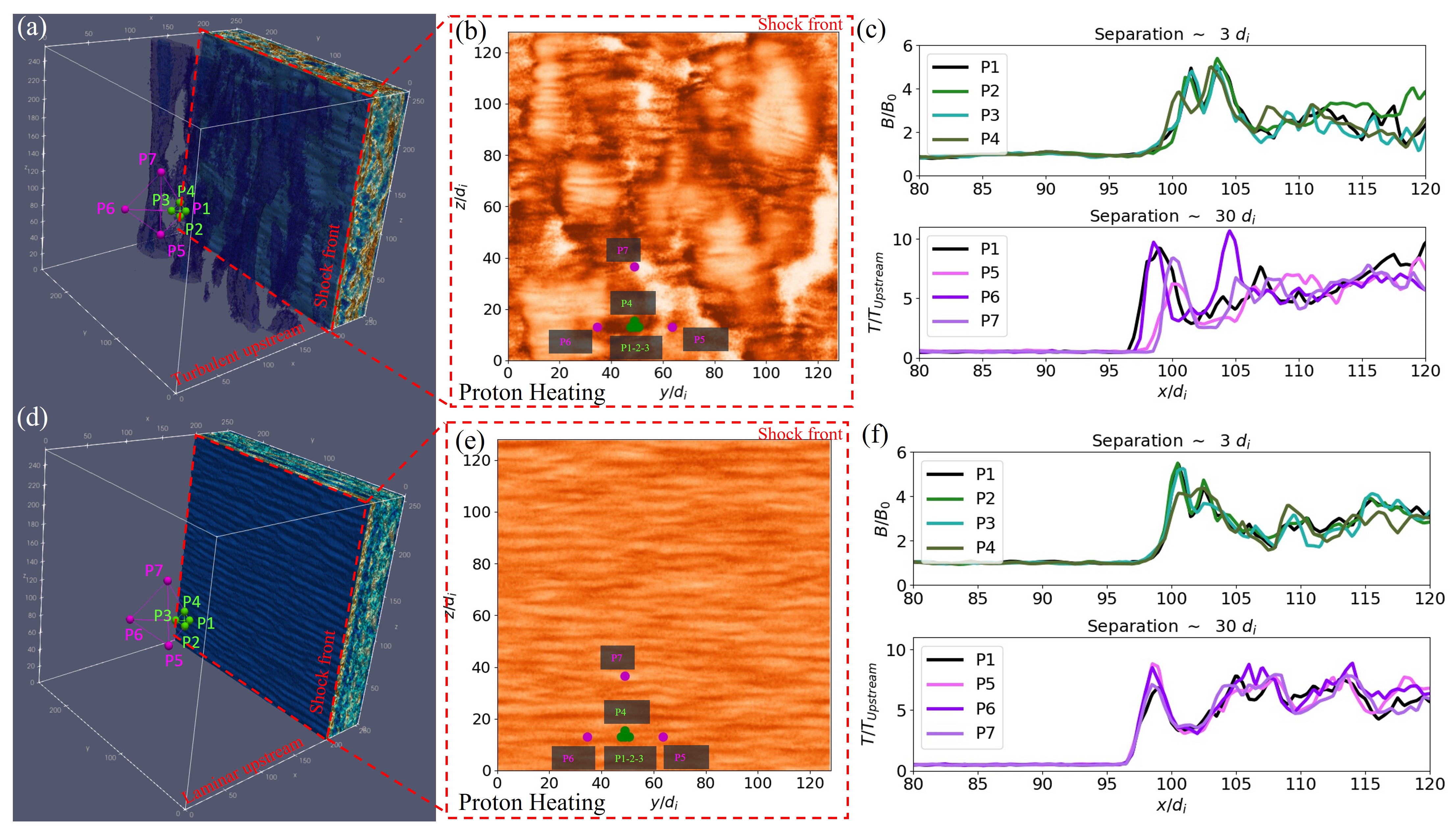}
    \caption{(a)-(d) Magnetic field rendering highlighting the shock front in the perturbed and laminar cases and the close upstream with seven virtual spacecraft arranged as two tetrahedra spaced at 3 and 30 $d_i$ (green and purple, respectively). (b)-(e) Color map showing proton heating $\mathrm{ T_p/T_{upstream}}$ along the shock front. (d)-(f) Virtual spacecraft observations along the shock normal direction for magnetic field and proton heating performed with the spacecraft tetraheadra spaced ad 3 and 30 $d_i$ (green and purple, respectively).}
    \label{fig:fig7_PO}
\end{figure*}

\subsection{Virtual spacecraft observations}
Numerical simulations are a crucial tool to advance our knowledge of spacecraft observations and assisting the design of new missions owing to the possibility of generating synthetic, virtual spacecraft measurements~\citep{Valentini2016,Perri2017,Pecora2023}. In this subsection, we discuss an example of such a study, applied to the interaction between shock and pre-existing turbulence. 

An emerging picture from the current work is that, when studying the shock propagation in a turbulent medium, shock ripples happening at the short wavelengths of $d_i$ are modulated by the turbulent fluctuations at larger scales, in a complex scenario for the shock front where different portions have different local geometry and environments. From a spacecraft measurements perspective, resolving simultaneously the short and long wavelength fluctuations present in cross-scale systems such as the one here described is extremely challenging. Such a challenge is inspiring new multi-scale, multi-spacecraft missions, such as HelioSwarm~\citep{Spence2019}, devoted to analyze plasma turbulence in the solar wind up to sub-ion scales, and Plasma Observatory \citep{Retino2022}, mostly focused on unveiling the fundamental mechanisms responsible for particle acceleration in the near-Earth environment including shocks and jets.

To this end, we elucidate what would be observed by the Plasma Observatory constellation in our simulation domain. In Figure~\ref{fig:fig7_PO} we show renderings of the computational domains with seven virtual spacecraft arranged as two tetrahedra sharing one vertex and at a separation of 3 (green) and 30 $d_i$ (purple), respectively few ion and fluid scales. For the purpose of these synthetic observations, we report the proxy for the proton heating, $\mathrm{T_p/T_{upstream}}$, along a two dimensional slice of the simulation domain showing the shock front (Panels (b)-(e)). The virtual spacecraft measurements of magnetic field at short separation show the difference in magnetic field increase observed due to shock rippling, as it can be seen comparing the P1-3 with the P4 plots in Figure~\ref{fig:fig7_PO}(c) and (g). The process of ion heating is highly structured at fluid scales, depending on several parameters such as the local magnetic field, which can be measured also at ion scales. While resolving the shock ripples, an important local property of the shock front, the tetraheadron with the larger spacing will resolve the larger scale shock front irregularity due to turbulence, as seen in the bottom panel Figure~\ref{fig:fig7_PO}(c).

Resolving such complex features of shock front variability would also be invaluable to advance our knowledge of particle acceleration at shocks. Indeed, with such multi-spacecraft measurements at different scales, it would be possible to understand which portions of the shock front are the most efficient at energising particles efficiently is achieved, distinguishing between processes such shock rippling operating at small scales and larger scales fluctuations possibly due to pre-existing turbulence, for example through the measurement of the departure from an average shock geometry as done for the simulations in Figure~\ref{fig:fig4_tbn}. This theme is extremely relevant for particle acceleration at the Earth's bow shock~\citep[e.g.,][]{Sundberg2016,Lindberg2022} as well as for other systems, such as interplanetary shocks~\citep[see][ for example]{Lario2008}.

\section{Conclusions}
\label{sec:conclusions}

In this work, we studied the interaction between supercritical, perpendicular shocks and fully developed, pre-existing plasma turbulence. We employ a novel simulation model, in which MHD and hybrid kinetic simulations are combined to obtain a collisionless shock wave propagating into an upstream characterised by fully developed turbulence. Our method builds onto previous studies in reduced dimensionality~\citep{Trotta2021,Trotta2022a}, and  is complementary to other methods looking at other interesting aspects of shock-turbulence interaction both in local configurations~\citep{Guo2012, Nakanotani2021,Nakanotani2022} and in global setups looking, for example, at planetary magnetospheres~\citep{Behar2022}.

The behaviour of a perpendicular, supercritical shock was studied in the unperturbed case and for two different levels of upstream turbulence, $\delta B/B_0 \sim 0, 0.5$ and 1, respectively. In the unperturbed case, shock rippling due to the perpendicular proton anisotropy driven by the reflected ion population is recovered, an important feature of perpendicular shocks, as studied in previous theoretical and numerical works~\citep{Hellinger1996,Burgess20163d}, and observed at the Earth's bow shock with closely-separated spacecraft constellations~\citep{Gingell2017, Johlander2018}. 

By coupling turbulent fields generated through compressible MHD simulations and hybrid kinetic simulations, for the first time in fully three-dimensional geometry, we addressed how turbulence is processed upon the shock crossing, with two interesting effects being observed: (i) increase in the level of fluctuations due to the compression at the shock, and (2) isotropisation of the magnetic field spectra in the close downstream. This may have important implications for the study of the nature of fluctuations associated with shock waves and their role in efficient particle acceleration, in particular for extra particle acceleration important in the shock downstream~\citep{Zank2015, Preisser2020b, Trotta2020b}. Further, interesting details of turbulence transmission across the shock, such as the study of the Yaglom law \citep{SorrisoValvo2019}, will be the object of a separate forthcoming work. Another important feature not studied here is the asymptotic behaviour of turbulence far downstream of the shock transition, for which simulations with larger domains and longer evolution times would be needed.

Concerning the shock transition when pre-existing turbulence is present, we discovered several interesting features. First of all, the shock front responds to upstream turbulence with corrugations following the turbulent field, an important feature that cannot be recovered considering only the fluctuations that are self-generated by the shock. In the moderately turbulent case $\delta B/B_0 \sim 0.5$, we still recover a rippled shock front, with ripples being modulated by the MHD-scale fluctuations. Such a behaviour may be important to understand the properties of shock accelerated particles interacting with such rippled portions of the shock front. 
For stronger perturbations, rippling becomes less prominent and the shock front is strongly distorted by the incoming turbulence. We found that, in the unperturbed case with the strongest rippled signature, the strongest local departures from the nominal shock geometry are achieved, with fluctuations happening over short $\sim d_i$ wavelengths, while in the perturbed case such departures from the nominal shock geometry are modulated over larger spatial scales. This has important implications with respect to observations, where such a variability may be important when looking at spacecraft crossing and inferring local shock parameters~\citep{Koval2008,Trotta2022b}.

To explain the variability in shock surface fluctuations and the different behaviour of shock rippling, we studied the proton temperature anisotropies in the simulations. We found that the presence of upstream turbulence introduces further complexity in the shock system, with the result of accelerating the processes restoring the equilibrium downstream of the shock. Analysis of the temperature anisotropy along the (perturbed) shock fronts is consistent with the picture of modified shock rippling in the presence of turbulence, suggesting a complex scenario for proton heating across shock waves.

This study has important implications on the theme of energy conversion at perpendicular shocks in various space and astrophysical settings where the role of pre-existing upstream turbulence is often neglected, though it is important to note that the scales simulated here are much smaller than those relevant in such systems, due to computational limitations. It is important to note that the behaviour of the shock rippling at short wavelength and the shock front corrugation due to turbulence happening at larger scales cannot be simultaneously resolved by closely spaced spacecraft constellations, motivating cross-scale missions of the future such as Plasma Observatory. Thus, our modelling effort provides important input for future missions design, constraining the required spacecraft constellations required to capture the complexity of the shock-turbulence interaction.

\section*{Acknowledgements}

This work has received funding from the European Unions Horizon 2020 research and innovation programme under grant agreement No. 101004159 (SERPENTINE, www.serpentine-h2020.eu).  Part of this work was performed using the DiRAC Data Intensive service at Leicester, operated by the University of Leicester IT Services, which forms part of the STFC DiRAC HPC Facility (www.dirac.ac.uk), under the project ``dp031 Turbulence, Shocks and Dissipation in Space Plasmas''. MHD simulations have been performed on the Newton HPC cluster at the University of Calabria, supported by ``Progetto STAR 2-PIR01 00008'' (Italian Ministry of University and Research). L.S.-V. is supported by the Swedish Research Council (VR) Research Grant N. 2022-03352. D.B. is supported by STFC grants ST/T00018X/1 and ST/X000974/1. H.H. is supported by the Royal Society University Research Fellowship URF\textbackslash R1\textbackslash180671. N.D.\ is grateful for support by the Academy of Finland (SHOCKSEE, grant No.\ 346902). L.P. is supported by the Austrian Science Fund (FWF): P 33285-N. XBC is supported by PAPIIT DGAPA grant IN110921.

\section*{Data Availability}
The simulation datasets used for the analyses in this work can be found and freely downloaded here: https://doi.org/10.5281/zenodo.7964045. The authors will share further datasets from the simulations upon request.



\bibliographystyle{mnras}
\bibliography{biblio} 








\bsp	
\label{lastpage}
\end{document}